\documentclass[12pt,a4paper,reqno]{article}

\usepackage{amsmath,latexsym,amssymb,amsthm}
\usepackage{graphicx}

\usepackage{amsmath}\usepackage{amssymb}
\usepackage{amsthm}

\newcommand{\be}{\begin{equation}}
\newcommand{\ee}{\end{equation}}
\renewcommand{\(}{\left(}
\renewcommand{\)}{\right)}

\renewcommand{\a}{\mathbf{a}}
\renewcommand{\b}{\mathbf{b}}

\renewcommand{\u}{\mathbf{u}}
\renewcommand{\v}{\mathbf{v}}
\newcommand{\w}{\mathbf{w}}
\newcommand{\x}{\mathbf{x}}
\newcommand{\y}{\mathbf{y}}
\newcommand{\z}{\mathbf{z}}
\newcommand{\del}{\nabla}

\renewcommand{\P}{\mathbf{P}}

\newcommand{\R}{\mathbb{R}}
\newcommand{\C}{\mathbb{C}}
\renewcommand{\H}{\mathcal{H}}

\newcommand{\ox}{\otimes}

\newcommand{\<}{\langle}
\renewcommand{\>}{\rangle}
\newcommand{\half}{\tfrac{1}{2}}

\newcommand{\transpose}{^{\mathrm{T}}}

\theoremstyle{plain} 
\newtheorem{theorem}{Theorem}[section]

\theoremstyle{definition}

\theoremstyle{remark}

\numberwithin{equation}{section}
\newcommand{\thmref}[1]{Theorem~\ref{#1}}

\begin{document}

\title{\bf{The geometric measure of multipartite entanglement and the singular values of a hypermatrix}}

\author{Joseph J. Hilling and Anthony Sudbery$^1$\\[10pt] \small Department of Mathematics,
University of York, \\[-2pt] \small Heslington, York, England YO10 5DD\\
\small $^1$ as2@york.ac.uk}

\date{24 June 2009}

\maketitle

\begin{abstract}
It is shown that the geometric measure of entanglement of a pure multipartite state satisfies a polynomial equation, generalising the characteristic equation of the matrix of coefficients of a bipartite state. The equation is solved for a class of three-qubit states.

\end{abstract}

\section{Introduction}

A pure state of a composite quantum system is entangled if it is not a product of states of the individual parts of the system. Thus one way of measuring the entanglement of the state (though there are many other independent types of entanglement \cite{Plenio-Virmani}) is to measure how different the state is from any product state. If we consider only pure states, this can be expressed in geometrical terms as the distance from the closest pure product state. The space of pure states of a system described by a Hilbert space $\H$ is the projective space $\P\H$ of one dimensional subspaces of $\H$, and the closeness of two pure states $\psi_1$ and $\psi_2$ is measured by the modulus of the inner product $|\<\psi_1|\psi_2\>|$ where $|\psi_1\>$ and $|\psi_2\>$ are normalised vectors representing the states; this gives rise to a standard metric, the Fubini-Study metric \cite{Wells:book}, on the projective space $\P\H$. In the multipartite situation the Hilbert space is a tensor product $\H = \H_1\ox\cdots\ox\H_n$, and the \emph{geometric measure of entanglement} \cite{Wei-Goldbart, Wei2004} or \emph{Groverian measure} \cite{Biham} of a pure state $|\Psi\>\in\H$ is a decreasing function of 
its Fubini-Study distance $g$ from the set of product states $|\phi_1\>\cdots |\phi_n\>$ with $|\phi_k\>\in\H_k$:
\[
g(|\Psi\>) = \max |\<\Psi|\(|\phi_1\>\cdots |\phi_n\>\)|^2
\]
where the maximum is taken over all single-system normalised state vectors $|\phi_1\>\in\H_1,\ldots ,|\phi_n\>\in\H_n$, and it is understood that $|\Psi\>$ is normalised.

The function $g$ is known as the \emph{injective tensor norm} on the tensor product space $\H_1\ox\cdots\ox\H_n$. In the bipartite case ($n = 2$), $g(|\Psi\>)$ is the largest coefficient in the Schmidt decomposition of $|\Psi\>$ (equivalently, the largest eigenvalue of the density operator $|\Psi\>\<\Psi|$, or the largest singular value of the matrix of coefficients of $|\Psi\>$ with respect to a product basis of $\H_1\ox\H_2$); for general $n$ it is the first coefficient in the multipartite generalisation of the Schmidt decomposition proposed in \cite{Schmidt}. It is determined by equations which constitute a nonlinear generalisation of the equations for the singular values and singular vectors of a matrix, the matrix being replaced by a ``hypermatrix'' (an $n$-dimensional array of numbers, here the coefficients of the given state with respect to a product basis). For these reasons, $g(|\Psi\>)$ is also known as the ``maximum eigenvalue" (more correctly, the ``maximum singular value") of the state $|\Psi\>$.

The definition of $g$ as a maximum makes it susceptible to numerical calculation, but it is not so easy to treat it analytically. It has been determined in certain special cases \cite{Wei-Goldbart, Wei2004, Shimoni, Tamaryan:geometric, Tamaryan:Groverian, Orus} but no general method of calculation has been proposed. In this paper we will give such a method, showing how the problem can be reduced in principle to the solution of a polynomial equation. We apply the method to three-qubit states. However, the polynomial equation in question has a very high degree, and even for three qubits the general case remains to some extent intractable. For a special class of three-qubit states, we obtain a complete solution, reproducing the results obtained by other authors \cite{Tamaryan:geometric} by methods which are specific to qubits.

The equations for the singular values and singular values of a hypermatrix are of mathematical interest in their own right. There does not seem to be a general theory available. We conclude by comparing the results for the particular case studied here with the familiar theorems for the matrix case. It appears that few of these theorems will have straightforward analogues in the general case.

\section{The nonlinear singular-value equations}

Given $|\Psi\>\in\H_1\ox\cdots\H_n$, we want to find $|\phi_k\>\in\H_k$, $k = 1,\ldots,n$, which maximise the function
\[
f(\phi_1,\ldots,\phi_n) = |\<\Psi|\(|\phi_1\>\cdots|\phi_n\>\)|^2
\]
subject to $\<\phi_k|\phi_k\> = 1$. Let $|e_i^{(r)}\>$ ($i = 1,\ldots,d_r = \dim\H_r$) be an orthonormal basis for $\H_r$ ($r = 1,\ldots,n$), and write 
\[
|\Psi\> = \sum_{i_1\ldots i_n}\overline{a}_{i_1\ldots i_n}|e_{i_1}^{(1)}\> \ldots |e_{i_n}^{(n)}\>, \qquad 
|\phi_r\> = \sum_{i_r}u_{i_r}^{(r)}|e_{i_r}^{(r)}\>.
\]
The problem becomes:
\[
\text{Maximise } f(\u) = \left|a_{i_1\ldots i_n}u_{i_1}^{(1)}\ldots u_{i_n}^{(n)}\right|^2\quad
\]
\[
\text{subject to }\quad \sum_{i_r}|u_{i_r}^{(r)}|^2 = 1 \quad (r = 1,\ldots n).
\]
where summation is understood over repeated lower indices $i_1,\ldots,i_n$. Introducing Lagrange multipliers $\lambda_r$, we are led to the equations \cite{Schmidt, Wei-Goldbart}
\begin{equation}
\label{multeigen}
a_{i_1\ldots i_n}u_{i_1}^{(1)}\ldots \widehat{u_{i_r}^{(r)}}\ldots u_{i_n}^{(n)} = \lambda_r\overline{u_{i_r}^{(r)}},
\end{equation}
where $\widehat{u_{i_r}^{(r)}}$ denotes that $u_{i_r}^{(r)}$ is to be omitted, and the overline denotes the complex conjugate. Multiplying by $u_{i_r}^{(r)}$ and using the constraints, we find that the $\lambda_r$ have a common value $\lambda$, which is the required extreme value of $f(\u)$. We will say that $\lambda$ is a \emph{singular value} of the hypermatrix $a_{i_1\ldots i_n}$, for the following reason.

In the simple case $n = 2$ the coefficients $a_{i_1 i_2}$ form a $d_1\times d_2$ matrix $A$, the unknowns $u_{i_1}^{(1)}$ and $u_{i_2}^{(2)}$ form a $d_1$-component vector $\u$ and a $d_2$-component vector $\v$, and the equations can be written
\begin{align}
A\v &= \lambda\overline{\u},\\
\u\transpose A &= \lambda \v^\dagger\\
\text{or}\hspace{1cm} A^\dagger\overline{\u} &= \lambda\v
\end{align}
where the dagger denotes the hermitian conjugate. Multiplying the first equation by $A^\dagger$ and using the second shows that $\v$ is an eigenvalue of $A^\dagger A$ with eigenvalue $\lambda^2$, i.e.\ $\lambda$ is a singular value of $A$. It satisfies the equation
\[
\det(A^\dagger A - \lambda^2 I) = 0.
\]
Our aim is to find the corresponding equation in the general case.
  
  If $\lambda = 0$ there is an established general theory for the equations \eqref{multeigen}. They can be written in terms of the multilinear function $\alpha$ of $n$ vector variables defined by 
\[
\alpha(\u^{(1)},\ldots,\u^{(n)}) = a_{i_1\ldots i_n}u_{i_1}^{(1)}\ldots u_{i_n}^{(n)}
\]
as
\begin{equation}
\label{singular}
\alpha = 0, \qquad \del_r\alpha = \mathbf{0} \quad (r = 1,\ldots,n) 
\end{equation}
where $\del_r$ denotes the gradient with respect to the vector variable $\u_r$. A solution $\u^{(1)},\ldots,\u^{(n)}$ of these equations is a \emph{critical point} of the function $\alpha$. The number of independent equations is one more than the number of variables, so the variables can be eliminated to give a polynomial equation in the coefficients $a_{i_1\ldots i_n}$:

\medskip
\emph{Theorem} (Cayley)\cite{Gelfand:book} \hspace{1em} There is a function hdet$(\alpha)$, polynomial in the coefficients $a_{i_1\ldots i_n}$, such that the equations \eqref{singular} have a solution with all $\u_i$ non-zero if and only if hdet$(\alpha) = 0$.

\medskip

The function hdet$(\alpha)$ is called the \emph{hyperdeterminant} of the hypermatrix $a_{i_1\ldots i_n}$. It is a special case of the \emph{discriminant} \cite{Gelfand:book} of a function of several variables: the discriminant $\Delta_f$ of a homogeneous polynomial function $f(x_1,\ldots,x_N)$ is a polynomial in the coefficients of $f$ with the property that $\Delta_f = 0$ if and only if there is a point $\x$ at which $f$ vanishes together with all its derivatives. The hyperdeterminant
has appeared elsewhere in the theory of multipartite entanglement \cite{Miyake:hyperdet, Miyake:classification}; for $n=3$ and $d_1 = d_2 = d_3 = 2$, its squared modulus is equal to the 3-tangle \cite{Woottang} of the 3-qubit state defined by the coefficients $a_{ijk}$. For $n = 2$ and $d_1 = d_2$, the hyperdeterminant reduces to the familiar determinant of the square matrix $a_{i_1i_2}$. In this case, as we have already noted, the determinant function also gives the condition for the equations \eqref{multeigen} to have a solution for any $\lambda$ (not just $\lambda = 0$); the characteristic equation of a square matrix, and the singular-value equation of a rectangular matrix, are both given by determinants. In general, however, the characteristic equation of a tensor is not so simply related to the hyperdeterminant function, and the theory needs to be extended.

\section{Generalising the Characteristic Polynomial}

In this section we shall derive a polynomial expression in the coefficients of a multipartite pure state, the roots of which include the geometric measure of entanglement of the state. This is directly analogous to the characteristic polynomial of a square matrix, which can be used to determine the Schmidt coefficients of an ordinary bipartite vector.

Our method is to consider the function $\alpha(\u^{(1)},\ldots,\u^{(n)})$ as a function of two of the variables with a matrix of coefficients which depend on the other variables. First we shall prove a lemma concerning such matrix-valued functions of several variables.

\begin{theorem}\label{lemma}
	Let $A(\z)$ be a $p\times q$ matrix function of $N$ complex variables $\z = (z_1,\ldots, z_N)$ which depends only on the variables $z_k$ and not on their complex conjugates $\overline{z}_k$, i.e the entries of $A(\z)$ are holomorphic functions of the $z_k$; let $\beta(\z)$ be a (non-holomorphic) real-valued function of $\z\in\C^N$; and let $\lambda$ be a positive real number. Then the real-valued function 
	$$\textrm{\emph{det}}[A(\z)^\dagger A(\z)-\lambda^2\beta(\z) I]$$
	vanishes along with its first derivatives with respect to the real and imaginary parts of $z_k$, at a point $\z$ with $\beta(\z) = 1$,  if and only if $\lambda$ is a singular value of $A(\z)$ and 
		\begin{equation} \label{Axy} 
		\x\transpose \frac{\partial A}{\partial z_k} \y = \lambda\frac{\partial\beta}{\partial z_k} 
	\end{equation}
	where $\x\in\C^p$ and $\y\in\C^q$ are respectively left and right singular vectors of $A(\z)$ corresponding to the singular value $\lambda$:
	\begin{align}\label{Rschm} A(\z)\y &= \lambda\overline{\x},\\
	\label{Lschm}  \x\transpose A(\z) &= \lambda\y^\dagger,\\
	\notag \x^\dagger \x = {}&\y^\dagger\y = 1.
	\end{align}.
	\end{theorem}
	
	Note: in eq. \eqref{Axy} $\beta$ is to be treated as a function of independent variables $\z$ and $\overline{\z}$ (i.e. $\displaystyle\frac{\partial\beta}{\partial z_k} = \frac{1}{2}\(\frac{\partial\beta}{\partial x_k} - i\frac{\partial\beta}{\partial y_k}\)$ where $z_k = x_k + iy_k$). 

\begin{proof}
Let the singular values of $A(\z)$ be $\sigma_1(\z),\ldots,\sigma_q(\z)$. Then we can write the determinant as
$$
P(\z,\lambda):=\det[A(\z)^\dagger A(\z)-\lambda^2\beta(\z)] =
\prod_i\(\sigma_i(\z)^2-\lambda^2\beta(\z)\).
$$ 
Calculating the derivative we have
$$
\frac{\partial P(\z,\lambda)}{\partial z_k}=\sum_i\left(\frac{\partial [\sigma_i^2]}{\partial z_k}-\lambda^2\frac{\partial\beta}{\partial z_k}\right)\prod_{j\neq i}\(\sigma_j(\z)^2-\lambda^2\)
$$
assuming $\beta(\z) = 1$. Now $P(\z,\lambda)=0$ is equivalent to $\sigma_i(\z)^2 = \lambda^2$, for some $i$, so $\sigma_i(\z) = \lambda$. Suppose $i=1$. If $\sigma_1(\z)^2$ is a degenerate eigenvalue of $A(\z)^\dagger A(\z)$ then the derivative will vanish, as the product term on the right hand side will always contain a zero term. Thus we shall assume that the eigenvalue is non-degenerate. In that case we have
$$
\frac{\partial P(\z,\lambda)}{\partial z_k}=\left(\frac{\partial [\sigma_1^2]}{\partial z_k}-\lambda^2 \frac{\partial \beta}{\partial z_k}\right)\prod_{j\neq 1}(\sigma_j^2-\lambda^2)
$$
The product term is non-zero, so 
\begin{equation}\label{equiv}
\frac{\partial P(\z,\lambda)}{\partial z_k}=0 \quad \Leftrightarrow \quad\frac{\partial [\sigma_1^2]}{\partial z_k} = \lambda^2 \frac{\partial\beta}{\partial z_k}.
\end{equation}

Now consider $\x$ and $\y$ as functions of $\z$ and $\overline{\z}$ defined by the singular-value equations
	\begin{align*}A(\z)\y &= \sigma_1(\z)\overline{\x},\\
  \x\transpose A(\z) &= \sigma_1(\z)\y^\dagger,\\
	\x^\dagger \x = {}&\y^\dagger\y = 1.
	\end{align*}
These give 
$$
\sigma_1^2 = \y^\dagger A^\dagger A\y,
$$
and therefore
\begin{align}
\frac{\partial [\sigma_1^2]}{\partial z_k}
&= \y^\dagger A^\dagger \frac{\partial A}{\partial z_k}\y 
+ \frac{\partial\y^\dagger}{\partial z_k}A^\dagger A\y + \y^\dagger A^\dagger A\frac{\partial\y}{\partial z_k}\\
&= \sigma_1\x\transpose\frac{\partial A}{\partial z_k}\y + \sigma_1^2\(\frac{\partial\y^\dagger}{\partial z_k}\y + \y^\dagger\frac{\partial\y}{\partial z_k}\)\\
&= \lambda\x\transpose\frac{\partial A}{\partial z_k}\y
\end{align}
since $\y^\dagger\y = 1$ is constant. The theorem now follows from \eqref{equiv} (since $P$ is real, ${\partial P}/{\partial z_k} = 0$ implies the vanishing of derivatives with respect to the real and imaginary parts of $z_k$).
\end{proof}

Now take $p = d_1$, $q = d_2$ and $N = d_3 + \cdots + d_n$, with $\x = \u^{(1)}$, $\y = \u^{(2)}$, $\z = (\u^{(3)},\ldots,\u^{(n)})$, and 
\begin{align*}
A(\z)_{i_1i_2} &= a_{i_1i_2i_3\ldots i_n}u_{i_3}^{(3)}\ldots u_{i_n}^{(n)},\notag\\
\beta(\z) &= \|\u^{(3)}\|^2\ldots\|\u^{(n)}\|^2
\end{align*}
where $\|\u\|^2 = \u^\dagger\u$. Then the equations \eqref{Axy}--\eqref{Lschm} become the nonlinear singular-value equations \eqref{multeigen}. Thus the nonlinear singular-value problem for $n$ vector variables has been reduced to the problem of finding critical points of a function of $n-2$ vector variables. The function 
$\det[A(\z)^\dagger A(\z)-\lambda^2\beta(\z) I]$ is a homogeneous polynomial function of the real and imaginary parts of the $n-2$ vector variables $\u^{(3)},\ldots\u^{(n)}$, so the existence of critical points is given by the vanishing of a discriminant, which is a polynomial in the coefficients of the function and therefore in $\lambda$. This polynomial function of $\lambda$ is the generalisation of the characteristic polynomial for a multilinear form. However, the vanishing of the discriminant does not guarantee that the critical points will be real, which we require since their coordinates are the real and imaginary parts of the original complex coordinates. Thus our characteristic equation is a necessary condition, but not a sufficient one, for the existence of singular values of the hypermatrix. In our study of three-qubit states we shall see examples of solutions of the characteristic equation which fail to be singular values.

To summarise:

\begin{theorem} \label{summary} Let $\alpha:\C^{d_1}\times\cdots\times\C^{d_n}\to\C$ be a multilinear function of $n$ vector variables:
\[
\alpha\(\u^{(1)},\ldots,\u^{(n)}\) = a_{i_1\ldots i_n}u_{i_1}^{(1)}\ldots u_{i_n}^{(n)}.
\]
For $\lambda\in\R$, define the real polynomial function $\tilde{\alpha}(\lambda):\R^{2d_3}\times\cdots\times\R^{2d_n}\to\R$ (identifying $\C^d$ with $\R^{2d}$) by 
\[
\tilde{\alpha}(\lambda)\(\u^{(3)},\ldots,\u^{(n)}\) = \det[A^\dagger A - \|\u^{(3)}\|^2\cdots\|\u^{(n)}\|^2 I]
\]
where the $d_1\times d_2$ matrix $A = A(\u^{(3)},\ldots, \u^{(n)})$ is defined by
\[
A_{i_1i_2} = a_{i_1i_2i_3\ldots i_n}u^{(3)}_{i_3}\ldots u_{i_n}^{(n)}.
\]
Suppose $\lambda \neq 0$. Then the equations
\begin{equation}
a_{i_1\ldots i_n}u_{i_1}^{(1)}\ldots \widehat{u_{i_r}^{(r)}}\ldots u_{i_n}^{(n)}= \lambda\overline{u_{i_r}^{(r)}}
\end{equation}
have a solution with all $\u^{(r)}$ non-zero if and only if $\tilde{\alpha}(\lambda)$ has a real critical point. If this is so, $\lambda$ satisfies the polynomial equation
\be\label{multichar}
\Delta_{\tilde{\alpha}(\lambda)} = 0.
\ee
\end{theorem}

\section{Three-qubit states}

We will now apply the general theory of the previous section to the simplest case after the familiar bipartite (matrix) case, namely the case of three qubits. Even here the characteristic equation of a general $2\times 2\times 2$ hypermatrix is dauntingly complicated. Instead of the most general three-qubit state, we will consider a state in the generalised Schmidt form
\be\label{Schmidt}
|\Psi\> = a|000\> + b|011\> + c|101\> + d|110\> + f|111\>
\ee
with $a,b,c,d$ real and positive, $a \ge b \ge c \ge d$, which has been shown \cite{Schmidt} to be a canonical form for three-qubit states.

This should be trivial. The first step in finding the canonical form for any state $|\Psi\>$ is to find its injective tensor norm $g(|\Psi\>)$; this gives the coefficient $a$. The remaining real coefficients $b,c,d$ are found by a further process of finding extrema of the same function whose maximum is $g(|\Psi\>)$. Thus we might expect that if $|\Psi\>$ is already in the form \eqref{Schmidt}, its characteristic equation should simply have the solutions $a,b,c,d$; the problem should be like diagonalising a matrix which is already diagonal. Interestingly, it is not so simple; so much so that we will need to simplify further by taking $f = 0$.

In this case the function $\tilde{\alpha}(\lambda)$ of \thmref{summary} is a function of a single vector variable $\z\in\C^2$, which we will regard as a function of the four variables $(z_1,z_2,\overline{z}_1,\overline{z}_2)$. It is a homogeneous quadratic in both $\z$ and $\overline{\z}$
and is therefore of the form
\[
\tilde{\alpha}_\lambda(\z,\overline{\z}) = t_{ijkl}z_i z_j\overline{z}_k\overline{z}_l
\]
where $t_{ijkl} = t_{jikl}$ and $t_{ijkl} = t_{ijlk}$. Suppose $\tilde{\alpha}_\lambda$ vanishes together with its derivatives at $\z = \a$, $\overline{\z} = \b$. Let $D_z\tilde{\alpha}_\lambda(\overline{\z})$ be the discriminant of $\tilde{\alpha}_\lambda$ regarded as a function of $\z$ alone; this function of $\z$ is singular if $\overline{\z} = \b$ (having the singularity $\z = \a$); thus $D_z\tilde{\alpha}_\lambda$ is a function of $\overline{\z}$ which has a zero at $\overline{\z} = \b$. We will now show that the derivatives of $D_z\tilde{\alpha}_\lambda$ also vanish at $\overline{\z} = \b$.

Since $\tilde{\alpha}_\lambda$ is a quadratic function of $\z$, its discriminant is the determinant of the $2\times 2$ matrix $u_{ij} = t_{ijkl}\overline{z}_k\overline{z}_l$:
\begin{align*}
D_z\tilde{\alpha}_\lambda &= \half\epsilon_{im}\epsilon_{jn}u_{ij}u_{mn}\\
&= \half\epsilon_{im}\epsilon_{jn}t_{ijkl}t_{mnpq}\overline{z}_k\overline{z}_l\overline{z}_p\overline{z}_q.
\end{align*}
Hence
\begin{align}\label{partialD}
\frac{\partial[D_z\tilde{\alpha}_\lambda]}{\partial\overline{z}_k}(\b)=2\epsilon_{im}\epsilon_{jn}t_{ijkl}t_{mnpq}b_l b_p b_q.
\end{align}
Now 
\[
\frac{\partial\tilde{\alpha}_\lambda}{\partial z_m}(\a,\b) = t_{mnpq}a_n b_p b_q = 0,
\]
so for $m = 1,2$ the vector $\v_m$ with components $(\v_m)_n = t_{mnpq}b_p b_q$ is orthogonal to $\a$ and therefore is a multiple of the vector with components $\epsilon_{ij}a_j$:
\[
t_{mnpq}b_p b_q = c_m\epsilon_{nr}a_r.
\]
By the symmetry of $t_{mnpq}$,
\[
c_m\epsilon_{nr}a_r = c_n\epsilon_{mr}a_r.
\]
It follows that $c_m$ is a scalar multiple of $\epsilon_{mr}a_r$, so
\[
t_{mnpq}b_p b_q = \sigma\epsilon_{mr}\epsilon_{ns}a_r a_s
\]
for some scalar $\sigma$. Now \eqref{partialD} gives
\[
\frac{\partial[D_z\tilde{\alpha}_\lambda]}{\partial\overline{z}_k}(\b) = t_{ijkl}a_i a_j b_l= \frac{\partial\tilde{\alpha}}{\partial\overline{z}_k} = 0.
\]
Thus $D_z\tilde{\alpha}_\lambda(\overline{\z})$ has a singularity at $\overline{\z}=\b$, and therefore its $\overline{\z}$-discriminant vanishes. This gives us the characteristic equation for $\lambda$.

Now we can find the equation for the injective tensor norm of the three-qubit state \eqref{Schmidt}. Thus we will apply \thmref{summary} to the function 
\[
\alpha(\x,\y,\z) = a_{ijk}x_iy_jz_k
\]
where $i,j,k = 0,1$ and the only non-zero values of $a_{ijk}$ are
\[
a_{000} = a, \quad a_{011} = b, \quad a_{101} = c, \quad a_{110} = d, \quad a_{111} = f.
\]
This notation is consistent with that of \thmref{lemma}, with $\z = \u^{(3)}$ being a single two-component vector.  The $2\times 2$ matrix $A(\z)$ is
\[
A(\z) = \begin{pmatrix}az_0&bz_1\\cz_1&dz_0 + fz_1\end{pmatrix},
\]
and the function $\tilde{\alpha}_\lambda$ is
\[
\tilde{\alpha}_\lambda(\z,\overline{\z}) = \det[A(\z)^\dagger A(\z) - \lambda^2\z^\dagger\z I].
\]
We write this as 
\[
\tilde{\alpha}_\lambda(\z,\overline{\z}) = F(\overline{z}_0,\overline{z}_1)z_0^2 + G(\overline{z}_0,\overline{z}_1)z_0 z_1 + H(\overline{z}_0,\overline{z}_1)z_1^2.
\]
Then
\[
D_z\tilde{\alpha}_\lambda(\overline{z}_0,\overline{z}_1) = G^2 - 4FH
\]
which is a homogeneous quartic function of $\overline{z}_0$ and $\overline{z}_1$:
\[
D_z\tilde{\alpha}_\lambda = \alpha(\lambda) \overline{z}_0^4 + \beta(\lambda) \overline{z}_0^3 \overline{z}_1 + \gamma(\lambda) \overline{z}_0^2 \overline{z}_1^2 + \delta(\lambda) \overline{z}_0 \overline{z}_1^3 + \epsilon(\lambda) \overline{z}_1^4,
\]
in which the coefficients $\alpha,\ldots,\epsilon$ are all quadratic functions of $\lambda^2$. The discriminant of this is the same as that of the associated function of one variable,
\[
P(x) = \alpha x^4 + \beta x^3 + \gamma x^2 + \delta x + \epsilon, 
\]
namely
\begin{align*}
\Delta &= 27B^2 - A^3\\[4pt]
\text{where }\qquad A &= \alpha\epsilon - \frac{\beta\delta}{4} + \frac{\gamma^2}{12},\\[4pt]
B &= \frac{\alpha\gamma\epsilon}{6} - \frac{\alpha\delta^2 + \beta^2\epsilon}{16} + \frac{\beta\gamma\delta}{48} - \frac{\gamma^3}{216}.
\end{align*}
Thus the characteristic equation $\Delta = 0$ has degree 12 in $\lambda^2$.

The equation simplifies considerably if we put $f = 0$. This special case has been investigated by Tamaryan et al.\ \cite{Tamaryan:geometric}, who have found some interesting solutions (in addition to $\lambda = a,b,c,d$) with an elegant geometric interpretation. It is of interest to see how these solutions emerge in our approach.

If $f = 0$ the coefficients $\beta$ and $\delta$ vanish, so that $D_z\tilde{\alpha}_\lambda(\overline{z}_0,\overline{z}_1)$ becomes a quadratic form in $(\overline{z}_0^2,\overline{z}_1^2)$, with coefficients
\begin{align*}
\alpha &= 4abcd(\lambda^2 - a^2)(\lambda^2 - d^2),\\
\gamma &= (\sigma_1^2 - 4\sigma_2)\lambda^4 + 4\sigma_3\lambda^2 - 8\sigma_4,\\
\epsilon &= 4abcd(\lambda^2 - b^2)(\lambda^2 - c^2),
\end{align*}
where $\sigma_k$ is the $k$th elementary symmetric function of $a^2,\,b^2,\,c^2$ and $d^2$. The discriminant of the quartic becomes
\begin{align*}
\Delta(\lambda) &= -\frac{\alpha\epsilon}{16}\(\gamma^2 - 4\alpha\epsilon\)^2\\[6pt]
&= -16a^2b^2c^2d^2(\lambda^2 - a^2)(\lambda^2 - b^2)(\lambda^2 - c^2)(\lambda^2 - d^2)
Q(\lambda)^2.
\end{align*}
Here $Q(\lambda)$ is a quartic in $\lambda^2$ which factorises completely:
\[
Q(\lambda) = 
\lambda^4[4S^2\lambda^2 - L^2][4(S^2 - abcd)\lambda^2 - (L^2 - 2abcd)]
\]
where $L$ and $S$ are the symmetric functions of $a,b,c,d$ introduced in \cite{Tamaryan:geometric}:
\begin{align*}
L^2 &= (ab + cd)(ac + bd)(ad + bc),\\
S^2 &= (s-a)(s-b)(s-c)(s-d),\\
s &= \half(a + b + c + d).
\end{align*}

The discriminant $\Delta(\lambda)$ has a zero of order 4 at $\lambda = 0$, but this is not covered by \thmref{summary}. In fact, if $\lambda = 0$ the equations \eqref{multeigen} become the equations for a singularity of the function $\alpha(\x,\y,\z)$, and the condition for a solution is the vanishing of the hyperdeterminant of $\a = (a_{ijk})$. But the known formula for a $2\times 2\times 2$ hyperdeterminant (\cite{Gelfand:book} p. 448) gives
\[
\text{hdet}(\a) = 4abcd,
\]
so $\lambda = 0$ is not a singular value.

This leaves as candidates for the injective tensor norm of the state
\[
|\Psi\> = a|000\> + b|011\> + c|101\> + d|110\>,
\]
i.e.\ the stationary
 values of the distance of the state from the set of separable pure states, or the singular values of the hypermatrix $a_{ijk}$,
\be\label{singvals}
\lambda = a,\; b,\; c,\; d,\; D = \frac{L}{2S} \text{ and } D' = \frac{L'}{2S'}
\ee 
where 
\begin{align*}
L'^2 &= L^2 - 2abcd = (cd - ab)(bd - ac)(bc - ad),\\
S'^2 &= S^2 - abcd\\ 
&= \tfrac{1}{16}(a+b+c+d)(a+b-c-d)(a-b+c-d)(-a+b+c-d).
\end{align*}
These are the solutions found by Tamaryan et al.\ \cite{Tamaryan:geometric}. It is not assumed that $S^2$, $S'^2$ or $L'^2$ are non-negative. Tamaryan et al.\ point out that if $S^2 \ge 0$, the positive real numbers $a,b,c,d$ satisfy quadrilateral inequalities (each of them is less than or equal to the sum of the others), and $S$ is the area of the cyclic quadrilateral of which they are the side-lengths, while $D$ is the diameter of its circumcircle. They also give a nice geometrical interpretation of $S'$ and $D'$: if $S'^2 \ge 0$, there is a self-intersecting cyclic quadrilateral $ABCD$ with side-lengths $a,b,c,d$ in which the vertices lie round the circle in the order $A,B,D,C$; the diameter of the circle is $D'$, and $S'$ can be interpreted in terms of signed areas.

In order to find the injective tensor norm we need to determine which of the solutions \eqref{singvals} is the largest. First we note that 
\[
L^2 - 4a^2 S^2 = \frac{a^2}{4}\(- a^2 + b^2 + c^2 + d^2 + \frac{2bcd}{a}\)^2 = \frac{r_a^2}{4} \ge 0
\]
where we follow the notation of Tamaryan et al.\ in defining $r_a$ (and $r_b$, $r_c$, $r_d$ similarly). From this we get the geometrically obvious result
\[ 
\frac{L^2}{4S^2} \ge a^2 \quad \text{if } S^2 > 0
\]
(no side of a cyclic quadrilateral can be greater than the diameter of the circumcircle), which also applies if $a$ is replaced by $b$, $c$ or $d$. Thus if the singular value $L/2S$ is real, it is at least as great as any of the values $a,b,c,d$.

Similarly,
\[
L'^2 - 4a^2 S'^2 = \frac{a^2}{4}\(- a^2 + b^2 + c^2 + d^2 - \frac{2bcd}{a}\)^2 = \frac{r_a^{\prime 2}}{4} \ge 0,
\]
from which it follows that 
\[ 
\frac{L'^2}{4S'^2} \ge a^2 \quad \text{if } S'^2 > 0.
\]
The singular value $L'/2S'$ can be real if $S'^2 < 0$, since $L'^2$ can be negative, but then the above inequality shows that it cannot be greater than any of $a,b,c$ or $d$.

Finally, to compare the sizes of the singular values $D = L/2S$ and $D' = L'/2S'$, we note that
\[
D'^2 - D^2 = \frac{1}{4S^2S'^2}[(L^2 - 2abcd)S^2 - L^2(S^2 - abcd)] = \frac{abcd}{4S^2S'^2}(L^2 - 2S^2)
\]
and we use the factorisation \cite{Tamaryan:geometric}
\[
L^2 - (a^2 + b^2 + c^2 + d^2)S^2 = \tfrac{1}{8}(a^2 + b^2 - c^2 - d^2)(a^2 - b^2 + c^2 - d^2)(a^2 - b^2 - c^2 + d^2).
\]
Since $a \ge b \ge c \ge d$, the first two factors are non-negative; hence if $S'^2 \ge 0$,
\[
D' \ge D \qquad \text{if and only if} \qquad a^2 + d^2 \ge b^2 + c^2.
\]

With this ordering of $a$, $b$, $c$ and $d$, the conditions for positivity of $S^2$ and $S'^2$ are:
\begin{align}
S^2 &\ge 0 \quad \Longleftrightarrow \quad a \le b + c + d,\\
S'^2 &\ge 0 \quad \Longleftrightarrow \quad a + d \le b + c.
\end{align}
Hence the greatest real solution of the characteristic equation, $\lambda_\text{max}\(|\Psi\>\)$, is

\[ 
\lambda_\text{max}(|\Psi\>) = \begin{cases} a &\text{ if } a \ge b + c + d,\\
D &\text{ if } a \le b + c + d \text{ but } a + d \ge b + c;\\
D &\text{ if } a + d \le b + c \text{ and } a^2 + d^2 \le b^2 + c^2;\\
D' &\text{ if } a + d \le b + c \text{ but } a^2 + d^2 \ge b^2 + c^2. 
\end{cases}
\]

However, this is not necessarily the injective tensor norm of $|\Psi\>$. For $\lambda$ to be a singular value of $a_{ijk}$, it is not sufficient that it satisfies the characteristic equation \eqref{multichar}; the associated singular vector has to satisfy reality conditions. If $\lambda$ satisfies the characteristic equation, we can find a singularity of the function $D_z\tilde{\alpha}_\lambda(\overline{\z})$ at some vector $\overline{\z}$. Fixing this value of $\overline{\z}$, and considering $\det[A(\z)^\dagger A(\z) - \lambda^2\z^\dagger\z I]$ as a function of $\z$, we are guaranteed to be able to find a singularity at some vector $\z$; but this may not be the complex conjugate of $\overline{\z}$.

If $\lambda = L/2S$ the function of $\overline{\z}$ becomes
\be\label{function}
D_z\tilde{\alpha}_\lambda(\z) = \frac{abcd}{64S^4}\(r_a^2r_d^2\,\overline{z}_0^2 -2r_a r_b r_c r_d\,\overline{z}_0\overline{z}_1 + r_b^2 r_c^2\,\overline{z}_1^2\)
\ee
whose non-zero singularities are given by 
\[
r_a r_d\, \overline{z}_0^2 - r_b r_c \,\overline{z}_1^2 = 0.
\]
Substituting for $z_1$ gives 
\[
\det[A(\z)^\dagger A(\overline{\z}) - \lambda^2 \z^\dagger\z I] = K\overline{z}_1^2\(r_a r_d\, z_0^2 - 2\sqrt{r_a r_b r_c r_d}\, z_0 z_1 + r_b r_c\, z_1^2\)
\]
where $K$ is a constant factor. This is singular at $\z$ where
\[
\sqrt{r_a r_d}\,z_0 - \sqrt{r_b r_c}\,z_1 = 0.
\]
Thus $z_0/z_1$ and $\overline{z}_0/\overline{z}_1$ are both equal to $\sqrt{r_b r_c/r_a r_d}$, which must be real if they are to be complex conjugates, so $r_b r_c/r_a r_d$ must be non-negative. With the ordering $a \ge b \ge c \ge d$, only $r_a$ can be negative, so the condition for $D$ to be a singular value of $a_{ijk}$ is $r_a \ge 0$, i.e.
\be\label{criterion}
b^2 + c^2 + d^2 + \frac{2bcd}{a} \ge a^2 \qquad \text{or} \qquad a^2 \le \frac{1}{2} + \frac{bcd}{a}.
\ee
We note that this cannot be satisfied if $a \ge b + c + d$, when $D$ is not real, so \eqref{criterion} is sufficient to determine whether the geometric measure is $a$ or $D$.

If $\lambda = L'/2S'$ the function $D_z\tilde{\alpha}_\lambda(\overline{\z})$ is given by \eqref{function} with $r_a,\ldots, r_d$ replaced by $r'_a,\ldots,r'_d$, and its non-zero singularities $\overline{\z}$ satisfy
\[
r'_a r'_d\, \overline{z}_0^2 - r'_b r'_c\, \overline{z}_1^2 = 0,
\]
but now substituting for $z_1$ gives 
\[
\det[A(\z)^\dagger A(\overline{\z}) - \lambda^2 \z^\dagger\z I] = K'\overline{z}_1^2\(r_a' r'_d\, z_0^2 + 2\sqrt{r'_a r'_b r'_c r'_d}\, z_0 z_1 + r'_b r'_c\, z_1^2\)
\]
so 
\[
\frac{z_0}{z_1} = \sqrt\frac{r'_b r'_c}{r'_a r'_d}, \qquad 
\frac{\overline{z}_0}{\overline{z}_1} = - \sqrt\frac{r'_b r'_c}{r'_a r'_d}.
\]
If these are to be complex conjugates of each other, they must both be imaginary; hence $r'_b r'_c/r'_a r'_d$ must be non-positive if $D'$ is to be a singular value of $a_{ijk}$.

But this cannot be true if $D'$ is the greatest solution of the characteristic equation, for then $a + d \le b + c$ but $a^2 + d^2 \ge b^2 + c^2$, so that $ad \le bc$; hence
\begin{align}
r_a' = - a^2 + b^2 + c^2 + d^2 - \frac{2bcd}{a} \le 2d^2 - \frac{2bcd}{a} \le 0,\\
r_b' = a^2 - b^2 + c^2 + d^2 - \frac{2acd}{b} \ge 2c^2 - \frac{2acd}{b} \ge 0,\\
r_c' = a^2 + b^2 - c^2 + d^2 - \frac{2abd}{c} \ge 2b^2 - \frac{2abd}{c} \ge 0,\\
\text{and }\qquad \qquad r_d' = a^2 + b^2 + c^2 - d^2 - \frac{2abc}{d} \le 2a^2 - \frac{2abc}{d} \le 0.
\end{align}
Therefore $D'$ cannot be the injective tensor norm of the state $|\Psi\>$.

Thus we reach the same conclusion as Tamaryan et al.: the injective tensor norm $g(|\Psi\>)$ of the normalised three-qubit state $|\Psi\> = 
a|000\> + b|011\> + c|101\> + d|110\>$ $(a \ge b \ge c \ge d)$ is determined by the single quantity $r_a = 1 - 2a^2 + 2bcd/a$:
\begin{align}
\text{If } r_a \le 0, &\text{ then } g\(|\Psi\>\) = a;\\
\text{if } r_a \ge 0, &\text{ then } g\(|\Psi\>\) = D = \sqrt{\frac{(s-a)(s-b)(s-c)(s-d)}{(ab+cd)(ac+bd)(ad+bc)}}.
\end{align}

\section{The singular vectors}

The singular vectors associated with the singular values  $a,b,c,d,D,D'$ (i.e.\ the solutions $(\u^{(1)}, \u^{(2)}, \u^{(3)}) = (\u,\v,\w)$ of the equations \eqref{multeigen}, or equivalently the product states $|\u\>|\v\>|\w\>$ having extremal overlap with $|\Psi\>$) are as follows.

For the singular values $a,b,c,d$ the results are as expected. In each case there is just one set of singular vectors (up to phase factors), corresponding to the three-qubit states $|000\>$, $|011\>$, $|101\>$, $110\>$. For the other two singular values $D$ and $D'$ the function $D_z\tilde{\alpha}_\lambda$ has two singularities, yielding either two sets of singular vectors or none. For $D$ singular vectors exist if $r_a$,$r_b$,$r_c$ and $r_d$ are all non-negative; these are \cite{Tamaryan:geometric}
\begin{align}\label{singvecs}
|\u\> &= \frac{\sqrt{r_c r_d}|0\> + \sqrt{r_a r_b}|1\>}{4S\sqrt{ab + cd}},\notag\\[6pt]
|\v\> &= \frac{\sqrt{r_b r_d}|0\> + \sqrt{r_a r_c}|1\>}{4S\sqrt{ac + bd}},\\[6pt]
|\w\> &= \frac{\sqrt{r_b r_c}|0\> + \sqrt{r_a r_d}|1\>}{4S\sqrt{ad + bc}}.\notag
\end{align}
Changing the sign of any of the square roots $\sqrt{r_a}, \sqrt{r_b}, \sqrt{r_c}, \sqrt{r_d}$ gives (up to phase factors) just one further set of vectors.

For the singular value $D'$ there are singular vectors if either one or three of $r'_a$, $r'_b$, $r'_c$, $r'_d$ are negative, so that one of them has a different sign from the others. Suppose the odd one out is $r'_a$; then the singular vectors are given by  
\begin{align}\label{singvecs2}
|\u\> &= \frac{\sqrt{r'_c r'_d}|0\> + \sqrt{- r'_a r'_b}|1\>}{4S\sqrt{cd - ab}},\notag\\[6pt]
|\v\> &= \frac{\sqrt{r'_b r'_d}|0\> + \sqrt{- r'_a r'_c}|1\>}{4S\sqrt{bd - ac}},\\[6pt]
|\w\> &= \frac{\sqrt{r'_b r'_c}|0\> + \sqrt{- r'_a r'_d}|1\>}{4S\sqrt{bc - ad}}.\notag
\end{align}
Again, changing any of the square roots gives one further set of vectors.

\section{Conclusion}

We have shown, in principle, how to solve the nonlinear singular-value equations which determine the generalised Schmidt coefficients of a pure multipartite quantum state (the singular values of the hypermatrix of coefficients of the state in a product bases). The largest of these Schmidt coefficients is the injective tensor norm, closely related to the geometric measure of entanglement. We have shown that it satisfies a polynomial equation which generalises the singular-value equation of a matrix, and we have shown explicitly how to find the characteristic polynomial for a three-qubit state (i.e. a $2\times 2\times 2$ hypermatrix). We have solved the equation to find the singular values and singular vectors for a class of such states.

It is interesting to compare our solutions with the bipartite case. The singular values and singular column vectors of an $m\times n$ matrix $A$ have the following properties:
\begin{itemize}
\item The singular vectors and null vectors of $A$ span $\C^n$.
\item The singular values are all real.
\item Every solution of the characteristic equation is a singular value.
\item The singular vectors associated with a particular singular value $\lambda$ form a vector subspace of $\C^n$, whose dimension is the multiplicity of $\lambda$ as a root of the characteristic polynomial.
\item Singular vectors associated with different singular values are orthogonal.
\end{itemize}

The example we have studied shows that most of these properties do not extend to the multipartite case (i.e.\ to higher-dimensional hypermatrices). The first property does survive in our example; indeed, in some cases there are enough singular vectors that not only do the singular vectors $|\u\>, |\v\>, |\w\>$ span the individual spaces, but their tensor products $|\u\>|\v\>|\w\>$ span the tensor product space. This is related to the failure of the the orthogonality property of singular vectors associated with different singular values: not only are they not orthogonal, they may not even be independent. Thus there may be more singular vectors than in the bipartite case. In another sense, there may be fewer singular vectors: those associated with a particular singular value form a discrete set (apart from phase factors), rather than a vector subspace. This is a feature of the nonlinearity of the problem. 

Our example suggests that something may survive of the link between the multiplicity of a singular value (as a root of the characteristic polynomial) and the number of independent singular vectors. However, there can be roots of the characteristic polynomial which are not singular values; they may not be real, or they may be real but have no singular vectors associated with them.

\section*{Acknowledgements}

We are grateful to Sayatnova Tamaryan for an extensive email correspondence, and for pointing out an error in an earlier version of this paper.


\end{document}